\documentclass[prl,twocolumn,showpacs,superscriptaddress]{revtex4-1}
\usepackage{graphicx,amssymb,amsmath,color,psfrag}
\begin{document}

\newcommand{\g}{}
\newcommand{\m}{\colod{red}}
\newcommand{\be}{\begin{equation}}
\newcommand{\ee}{\end{equation}}
\definecolor{darkgreen}{rgb}{0,0.6,0}

\title{Generalized Gibbs ensemble and  work statistics of a quenched Luttinger liquid}

\author{Bal\'azs D\'ora}
\email{dora@kapica.phy.bme.hu}
\affiliation{BME-MTA Exotic  Quantum  Phases Research Group, Budapest University of Technology and
  Economics, Budapest, Hungary}
\affiliation{Department of Physics, Budapest University of Technology and Economics, Budapest, Hungary}
\author{\'Ad\'am B\'acsi}
\affiliation{Department of Physics, Budapest University of Technology and Economics, Budapest, Hungary}
\author{Gergely Zar\'and}
\affiliation{BME-MTA Exotic  Quantum  Phases Research Group, Budapest University of Technology and
  Economics, Budapest, Hungary}

\date{\today}

\begin{abstract}
{ We analyze the probability distribution function (PDF) of work done
on a Luttinger liquid for an arbitrary finite duration interaction 
quench and show that it can be described in terms a
 generalized Gibbs ensemble. We construct the corresponding 
density matrix with explicit intermode correlations, and determine the
duration and interaction dependence of the probability of an adiabatic
transition and the PDF of non-adiabatic processes.}
In the thermodynamic limit,  the PDF of work exhibits 
a { non-Gaussian} maximum around the excess heat, carrying 
almost all spectral weight. In contrast, in the small system limit 
most  spectral weight is carried by a delta peak
at the energy of the adiabatic process, and
 an oscillating PDF with dips at
energies commensurate to the quench duration and with an exponential  
envelope develops. Relevance to cold atom experiments is also discussed.
\end{abstract}

\pacs{05.30.Jp,71.10.Pm,05.70.Ln,67.85.-d}

\maketitle

{\em Introduction.} 
Non-equilibrium many-body dynamics constitutes a terra incognita in
comparison to its equilibrium counterpart. 
Its exploration has begun recently by a series of experiments on cold
atomic
gases\cite{greiner1,kinoshita,hofferberth,BlochDalibardZwerger_RMP08}
and other systems\cite{utsumi}, triggering  
valuable theoretical works\cite{dziarmagareview,polkovnikovrmp}. A
number of interesting issues has been analyzed, such as thermalization  
and equilibration and their relation to integrability,
defect and entropy production due to universal near adiabatic
dynamics, quantum fluctuation relations\cite{rmptalkner}, non-linear
response etc.

Monitoring non-adiabatic dynamics provides a great deal of information
about the universal features of the quantum system at hand. The
scaling of expectation values or  the first few moments of observables (e.g 
the defect density)  after a quench through a quantum critical point can be expressed in terms of the equilibrium critical 
exponents\cite{dziarmagareview,polkovnikovrmp}.
However, the full characterization of a quantum state is only possible through its all higher moments, encoding unique information about 
non-local correlations of arbitrary order  and entanglement\cite{gritsev}.
This is equivalent to determining the full distribution function of the quantity of interest.
While its equilibrium evaluation is already rather involved\cite{gritsev}, obtaining the full non-equilibrium 
distribution function of a physical observable has rarely been carried
out\cite{gring}. 

A delightful exception is the statistics of work done during a quench,
which has been studied in Refs. \cite{silva,venuti1} for a \emph{sudden quench} 
between \emph{gapped} phases, separated by a quantum \emph{critical
  point} (and gap closing). 
The probability distribution function (PDF) of work done, $P(W)$, 
 involves all possible  moments of energy\cite{rmptalkner},  
thus providing us with full characterization of the energy distribution.

{ While the transition between two 
gapped phases is of great interest, many interacting 
one-dimensional systems form  gapless
Luttinger liquid (LL) states~\cite{giamarchi}.} 
In particular, interacting cold atoms in a one dimensional trap, { e.g.,} 
often form such LL's,  as also confirmed by
experiments~\cite{kinoshita,hofferberth,haller,hofferberthnatphys,paredes},
but LL states appear in various spin models or interacting fermion 
systems~\cite{giamarchi}.
 This state of matter is characterized by bosonic
collective modes as elementary excitations, 
{ and  by especially strong quantum fluctuations.}
 How this system reacts to a time dependent protocol, i.e. a quantum
 quench, is a highly nontrivial problem, though some of its properties
 have already 
been analyzed\cite{cazalillaprl,iucci,uhrig,doraquench,perfettoEPL}. 

{ Here we shall study the PDF of work on this 
 prototypical example of a Luttinger liquid,  determine 
$P(W)$ after an {\em arbitrary quench protocol}, and  also
  construct explicitly the generalized Gibbs ensemble
 which reproduces all  moments of $P(W)$.
We remark that this is one of the rare occasions, where the 
generalized Gibbs ensemble can be constructed analytically for an
interacting model.}
{ The study of an arbitrary quench protocol is inspired by the observation
that}, in reality, quenches are neither completely adiabatic nor
instantaneous { and, --- as we demonstrate through the properties of 
$P(W)$, --- } the characteristic  quench time is a crucial parameter 
of the quench itself. 
The work PDF is found to exhibit several universal forms (Gumbel or exponential distribution, e.g.),  as controlled by the system size and 
interaction dependent many-body orthogonality exponent, $\alpha$,  and the duration of the quench.

{\em Hamiltonian.}
{ We consider }an inherently gapless  system 
of hard core bosons (or an initially non-interacting Fermi gas) 
in one dimension,  which is interaction quenched by a 
given protocol into a final LL liquid state.
The corresponding LL Hamiltonian reads~\cite{doraquench,giamarchi} 
\begin{equation}
H=\sum_{q\neq 0} \omega_q(t)b_q^+ b_q
+\frac{g_q(t)}{2}[b_qb_{-q}+b_q^+b_{-q}^+]\;.
\label{hamilton}
\end{equation}
Here $\omega_q(t)=v(t)|q|$, and $v(t)=v+\delta v\; Q(t)$, with 
$v$ the bare "sound velocity",  $\delta v$
 its renormalization arising from interaction, 
and $b_q^+$ the creation operator of a bosonic density wave. 
The interaction $g_q(t)=g_2(q)|q|\,Q(t)$ and the velocity are changed 
within a quench time $\tau$, 
 with the  quench protocol $Q(t)$  satisfying
$Q(t<0)=0$ and $Q(\tau < t)= 1$.
For a linear quench,  in particular, $Q(0<t<\tau)=t/\tau$. 
Eq. \eqref{hamilton} constitutes the effective model for bosons
quenched away from the hard-core limit as well as 
 for fermions quenched away from the non-interacting limit, or for an 
XXZ spin chain~\cite{doraquench,giamarchi}, {\g though our findings apply to interacting initial states as well\cite{EPAPS}.}

{ Since Eq.~\eqref{hamilton} is quadratic, the time evolution 
can be formally determined exactly.}
From the Heisenberg equation of motion, we obtain~\cite{doraquench}
\begin{gather}
b_q(t)= u_q(t)\;b_q(0) +v^*_q(t)\;b^+_{-q}(0)\;,
\label{timedepop}
\end{gather}
where the time dependence is carried by the time dependent Bogoliubov
coefficients  
 $u_q(t)$ and $v_q(t)$, satisfying
\begin{gather}
i\partial_t\left[\begin{array}{c}
u_q(t)\\
v_q(t)\end{array}\right]=\left[\begin{array}{cc}
\omega_q(t) & g_q(t)\\
-g_q(t) & -\omega_q(t)
\end{array}\right]
\left[\begin{array}{c}
u_q(t)\\
v_q(t)\end{array}\right],
\label{beq}
\end{gather}
with the initial condition $u_q(0)=1$, $v_q(0)=0$.
%

{\em Generating function of work. }
Armed with the formal solution of the time-dependent Bogoliubov
equations, Eq.~\eqref{timedepop}, we analyze the statistics of work done.
Albeit the work done has been studied in classical statistical
mechanics exhaustively, its quantum generalization has been
carried out only recently~\cite{rmptalkner}, { and its properties are
known for very few systems}.
The quantum work cannot be represented by a single
Hermitian operator 
($\Leftrightarrow$ not an observable), but rather its characterization
requires two 
successive energy measurements, one before and one after the time dependent protocol (thus work characterizes a process).
The knowledge of all possible outcomes of such measurements yields the full probability distribution function (PDF) of work done on the system.

The characteristic function of work after the quench, 
{ $G(\lambda)\equiv \int dW\, e^{i W \lambda}\, P(W)$ can be expressed}
as~\cite{rmptalkner}
\begin{gather}
G(\lambda,\tau)=\langle \exp[i \lambda
  H_H(t>\tau)]\exp[-i\lambda H_H(0)]\rangle\;,
\label{eq:generator}
\end{gather}
where $H_H(t)$ 
is the Hamilton in the Heisenberg picture, 
and the expectation value is taken with the initial thermal state.
{ For a sudden quench (SQ),  $\tau=0$, and 
$G(\lambda,\tau)$
coincides with the Loschmidt echo~\cite{silva}. }
The expectation value, Eq. \eqref{eq:generator}
is independent of  $t$ for $t>\tau$, but depends on the quench { protocol
and its} duration $\tau$. 
$H_H(t)$ is obtained by expressing  the time dependent boson operators 
in Eq. \eqref{hamilton} using Eq.  \eqref{timedepop}. 
{ Eq.~\eqref{eq:generator} can then be evaluated at $T=0$ temperature
using identities familiar from the theory of squeezing operators,} 
yielding~\cite{EPAPS}
\begin{gather}
\ln\; G(\lambda,\tau)=i\lambda E_{ad}
-\sum_{q>0}\ln\left(1+n_q(1-e^{2i\,\Omega_q\lambda})\right),
\label{charworknq}
\end{gather}
with $E_{ad}= E_f-E_i$ the difference between the adiabatic ground state
energies in the final and initial state, and
$n_q={[\omega_q(t)-\Omega_q+2\textmd{Im}
\{v^*_q(t)\partial_t v_q(t)\} ]}/{2\Omega_q}$
the occupation number of mode $q$  in the final LL state, and  
$\Omega_q=\sqrt{\omega_q^2(t>\tau)-g^2_q(t>\tau)}$  
{ the corresponding excitation energy}~\cite{giamarchi}.

{\em Generalized Gibbs ensemble.}
The fact that
 Eq. \eqref{charworknq} depends only on the occupation numbers of
the steady state  indicates that a
generalized Gibbs ensemble (GGE)  
{ may describe the final state}~\cite{polkovnikovrmp}. 
{ The analytic construction of the final density matrix 
is usually an inadmissible task. Therefore, one typically 
focuses only on few body observables, and tries to build an
approximate density matrix describing these. 
Such an approach is, however, unable to account for 
the complete PDF of  work, which
depends on all possible moments of energy.}

In our case, the  final Hamiltonian can be diagonalized
by a Bogoliubov transformation
giving $H_f=\sum_{q\neq 0}\Omega_q
\hat n_q + E_f$.  In the steady state ($t\gg\tau$),  
the $\hat n_q$'s and their arbitrary products are 
constants of motion, and therefore 
the density matrix of the GGE should be built up, in principle, 
 from all of these  operators~\cite{iucci}. 
{ We find, however,  that for a $T=0$ temperature quench the 
density operator  
\begin{gather}
\hat\rho_{G}=\frac{1}{Z_{G}}\prod_{q>0}\exp\left[-\beta_q\Omega_q\hat n_q\right]\delta_{\hat n_q,\hat n_{-q}}\;,
\label{rhoGGE}
\end{gather}
accounts for all intermode correlations in the final state.
Here the mode dependent inverse temperatures 
$\beta_q$ are defined through $n_q\equiv \langle\hat n_q\rangle
\equiv1/[\exp(\beta_q\Omega_q)-1]$, and 
$Z_G=\textmd{Tr}\{\exp[-\sum_{q>0}\beta_q\Omega_q\hat n_q]\}$.
Indeed, it is easy to show that 
${\rm Tr} \{\hat\rho_{G}\,e^{i\lambda (H_f-E_i)}\}$   
reproduces $G(\lambda,\tau)$ and thus  the complete work distribution\cite{EPAPS}.  
Moreover, it gives back the expectation
value of any operators in the steady state.
Notice that the delta-functions in $\hat\rho_{G}$ 
imply  perfect correlations between the 
mode pairs $\pm q$.}
   
The structure of Eq. \eqref{rhoGGE}
 follows from the observation that, while time evolution 
does not conserve the number of bosons in a given pair of modes 
$\pm q$, it preserves $\hat n_q(t) -\hat n_{-q}(t)$.  
Since  the  only non-zero element 
of the initial density matrix corresponds to $\hat n_q=\hat n_{-q}=0$ 
at zero temperature,  this can only evolve along the 
diagonal "direction", $\hat n_q(t)-\hat n_{-q}(t)=0$.
{ The assumption  that evolution during the 
 quantum quench thermalizes the energy distribution of 
a given momentum pair with this constraint then  
amounts in the density matrix, Eq.~\eqref{rhoGGE}.
A given pair of modes thus thermalizes only along the diagonal of the
density matrix,   $\hat n_q=\hat n_{-q}$,  characterized by an
 effective inverse temperature $\beta_q$, while 
the weight of the non-diagonal  states 
 $\hat n_q\ne \hat n_{-q}$ remains zero, as in the initial state. 
Though umklapp processes may lead to further thermalization 
at larger time scales,} this structure is expected to be
stable within experimental time scales~\cite{EPAPS}.

%

{\em Perturbative generating function.}
{ Though an exact solution is formally also possible, 
the general properties of the final work PDF are already captured
  by} a more transparent perturbative solution of 
Eq.~\eqref{beq}~\cite{doraquench}. 
{ We thus}
expand Eq. \eqref{charworknq} for small $g_2(q)$ and $\delta v$,
and get { for large system sizes $L$}
%
\begin{gather} 
\ln  G(\lambda,\tau) =
iE_{ad}\Bigl( \lambda-\int\limits_0^\tau\int\limits_0^\tau dt_1 dt_2
Q'(t_1)Q'(t_2)\tau_0 \times 
\nonumber
\\
\times \left[ f(t_1-t_2+\lambda)- f(t_1-t_2)\right] \Bigr)\;.
\label{workgenquench}
\end{gather}
Here $E_{ad}=-({L/v})({g_2}/{v\tau_0})^2/16 \pi +\dots<0$ and 
$ f(t)=\tau_0/(t+i\tau_0)$,  
with $\tau_0$ a short time  cut-off associated with the finite range
of interaction,  $g_2(q)=g_2 \exp(-\tau_0 v |q|)$.  
Interestingly, the velocity renormalization, $\delta v$ does not enter  to lowest order. 
The cumulants, $C_n$ of the work done
can be derived by expanding Eq. \eqref{workgenquench} in $\lambda$
(see \cite{EPAPS}).

{\em Work  PDF: generic properties.}
{To analyze the PDF of work it is worth introducing the {\em
    dimensionless work}, measured with respect to the adiabatic ground
  state energy shift,} 
\be 
w\equiv (W-E_{ad})/{|E_{ad}|}\;.
\ee
{ The distribution of $w$ is then 
obtained by Fourier transforming $G(\lambda,\tau)$ as} 
\begin{gather}
p( w)=
{\mathcal P}_{ad}\;\delta(w)+\rho(w)\;.
\label{p(w)}
\end{gather}
The Dirac-delta peak corresponds to the probability 
of staying in the   adiabatic ground 
state, while the broad structure $\rho(w)$ is associated with 
transitions to excited states with $w>0$.  
The weight $\mathcal P_{ad}$ can be expressed as
\begin{gather}
\ln\left(\mathcal P_{ad} \right)=-i\alpha \int\limits_0^\tau \int\limits_0^\tau dt_1 dt_2 {Q'(t_1)Q'(t_2)} f(t_1-t_2)\;.
\label{Pad}
\end{gather}
The prefactor  $\alpha=|E_{ad}\tau_0|\sim N(g_2/v)^2$  denotes 
the total angle of
Bogoliubov rotations ($N\sim L/v \tau_0$ is the number of particles), and
can be viewed as  the many-body orthogonality  exponent. It 
 is also closely related to  the fidelity susceptibility~\cite{rams}. 
Alternatively, we can rewrite it as  $\alpha \sim L/\textit{l}$ 
with $\textit{l}$ the mean free path. Thus $\alpha \gtrless 1$ 
describes, using fidelity nomenclature, the thermodynamic / small 
system limits~\cite{rams} or, alternatively,  
 corresponds to the diffusive/ballistic limits, respectively, 
depending on the picture used. 

\begin{figure*}[t!]

\hspace*{4mm} $\alpha=20$ (thermodynamic limit) \hspace*{16mm} $\alpha=4$ (crossover region) \hspace*{1.7cm} $\alpha=0.2$ (small system limit)
\vspace*{0.4mm}
\includegraphics[width=5.5cm]{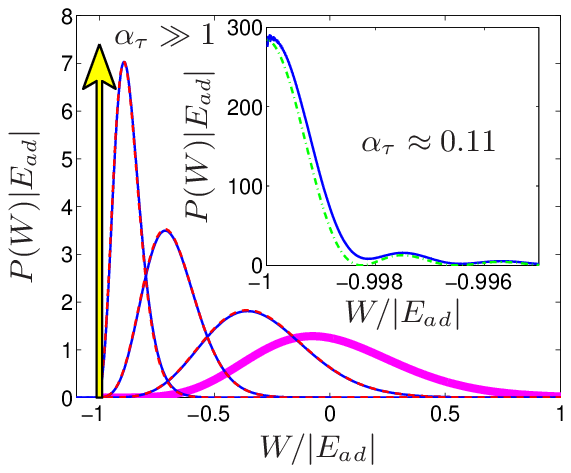}
\hspace*{1mm}
\includegraphics[width=5.5cm]{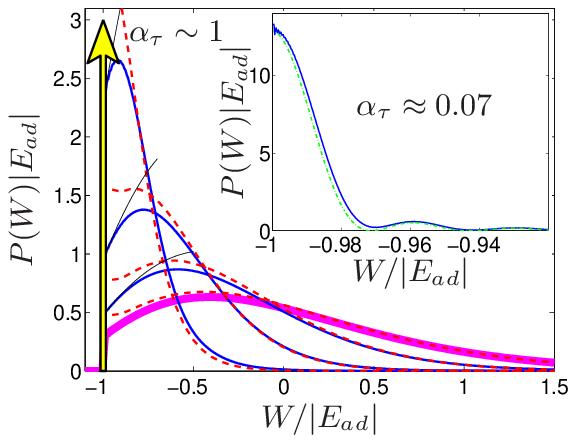}
\includegraphics[width=5.5cm]{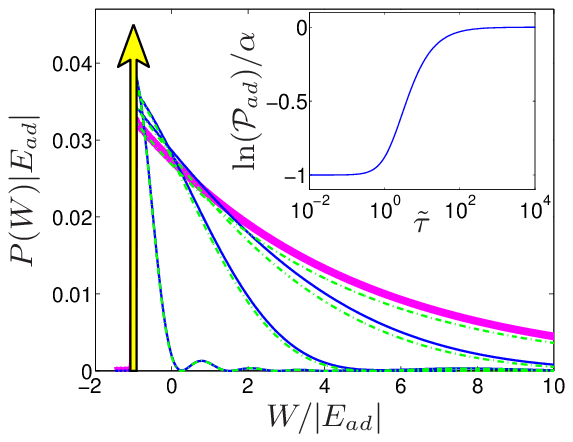}
\caption{(Color online) The PDF of work done on a LL is plotted after
  a linear quench from the numerical evaluation of
  Eq. \eqref{workgenquench} (blue solid line).  
Left panel: $\alpha=20$  with $\tilde\tau=0$, 
1, 2.5 and 5 from right to left and 180 (inset, $ P(W>E_{ad})$ only); 
middle panel: $\alpha=4$ with $\tilde\tau=0$, 1, 2 and 4 with increasing peak height and 55 (inset); right panel: 
$\alpha=0.2$ with $\tilde\tau=0$, 2, 5, and 25   from right to left.
The thick magenta line denotes the exact  SQ expression (Eq. \eqref{pdfSQ}), the red dashed line represent Eq. \eqref{pdfsteepest},
the thin black line in the middle panel visualizes Eq. \eqref{smallw}, while the green dash-dotted line 
shows Eq. \eqref{smallgamma}.
The vertical arrow at $W=E_{ad}$ denotes the Dirac-delta peak, whose spectral weight $\mathcal P_{ad}$ is
shown in the inset of the right  panel on semilog scale as a function of the ramp time $\tau$. 
\label{pdfworktau}}
\end{figure*}

In the adiabatic limit ($\tau\rightarrow \infty$), { a finite} system always
stays in its ground state, and the time evolved wave function coincides
with   the lowest energy eigenfunction of the instantaneous Schr\"odinger
equation~\cite{pollmann}. Consequently, only the first term remains in
Eq. \eqref{p(w)} with $\mathcal P_{ad}=1$. 
For  $\tau\ll\tau_0$, on the other hand, $\mathcal P_{ad}$  scales as
$\sim\exp(-\alpha)\sim \exp(-cst. \;L)$ (see Fig. \ref{pdfworktau}),
and in the limit  $L\rightarrow\infty$ --- but fixed interaction  --- 
$\mathcal P_{ad}$ vanishes due to the orthogonality catastrophe.

{\em Sudden quench (SQ) limit.}
In the extreme limit of a SQ, 
$\tau\ll \tau_0$, 
$G(\lambda,\tau)$ simplifies to 
$G(\lambda)=\exp\left[{iE_{ad}\lambda^2}/{(\lambda+i\tau_0)}\right]$,
and the  continuum part of the PDF of work is evaluated exactly as
\begin{gather}
\rho_{\rm SQ}(w)=\mathcal P_{ad}\;\exp(-\alpha w)\;{\alpha}\;{w^{-1/2}}\;
 I_1\bigl(2\alpha\sqrt {w}\bigr)\; ,
\label{pdfSQ}
\end{gather}
with $\mathcal P_{ad}=\exp(-\alpha)$
and $I_1(x)$  the modified Bessel function of the first kind. 
This is  the non-central $\chi^2$ distribution with 
non-centrality parameter $4\alpha$ in the limit of zero degrees of freedom~\cite{siegel}.
The average work is zero~\cite{doraquench}, { since for a SQ the
  system remains in its initial state and --- on average --- there is
  no back  reaction. Entropy is,  however,} 
generated by populating high and low energy configurations. 

The shape of $\rho(w)$ depends crucially on the orthogonality
parameter, $\alpha$. In the thermodynamic limit, $\alpha\gg 1$, 
almost all probability  weight is carried by a  peak  centered 
at around $W=0$ ($w=1$) and of  width $\Delta W\sim |E_{ad}|/ \sqrt{\alpha}$,
\begin{gather}
\rho_{\rm SQ}^{\alpha\gg 1}(w\gg \alpha^{-2})\approx \dfrac{\exp\left(-\alpha \left[1-\sqrt{w}\right]^2\right)}{w^{3/4}\sqrt{4\pi\alpha^{-1}}}\;,
\label{gaussSQ}
\end{gather}
whose high energy tail decays according to the  Gamma distribution,
 $\sim \exp(-\alpha  w)/w^{3/4}$.
In the small system regime $\alpha\ll 1$, { on the other hand,} the delta function retains almost all 
weight, 
and transfers only a fraction $\sim \alpha$ 
to an exponential distribution of 
 width $\Delta W\sim |E_{ad}|/\alpha $ and threshold at $E_{ad}$
for $w\ll\alpha^{-2}$.
In { the cross-over regime}, $\alpha \sim 1$, the maximum shifts 
to lower energies and the PDF of work develops a sizable value right
above the threshold at $E_{ad}$   (see Fig. \ref{pdfworktau}). 
The maximum of $P(W)$ occurs at $W>E_{ad}$  for
$\alpha>2$, while the PDF becomes monotonically decreasing for $\alpha<
2$. 

{\em Finite quench times.}
{ For finite  duration quenches, 
in addition to the orthogonality parameter $\alpha$, 
the work statistics also depends on $\tau$ and the
protocol $Q(t)$ itself.  For definiteness, we focus here 
on a linear quench~\cite{EPAPS}, and measure the degree of
adiabaticity by $\tilde  \tau=\tau/\tau_0$.}

{ For a finite duration quench, $\tilde\tau >1$, 
only a fraction $1/\tilde \tau$ of the
excitations experiences the quench as sudden. Consequently, 
in the expression of ${\cal P}_{ad}$, the orthogonality 
exponent $\alpha$ is replaced by  $\alpha_\tau\sim \alpha/\tilde\tau$,
and  ${\cal P}_{ad}$ becomes a monotonously increasing function 
of $\tilde \tau$ (see Fig.~\ref{pdfworktau}).
The crossover with increasing  $\alpha_\tau$ from $\mathcal P_{ad}\lesssim 1$ to vanishingly small spectral weight, $\mathcal P_{ad}$, occurs 
at $\alpha\sim \tilde\tau$.

}

{ Close to the threshold},  $W-E_{ad} \ll 1/\tau$, only 
states with energy smaller than $1/\tau$ and thus feeling a SQ 
contribute to work. Therefore, 
apart from a normalization factor, the PDF of work  agrees
with the SQ result, 
\begin{gather}
\rho(w\ll \alpha^{-1}\tilde\tau^{-1})\approx \mathcal P_{ad} \exp(\alpha) \rho_{SQ}(w)
\;,
\label{smallw}
\end{gather}
{ and  depends} on $\tau$ only through $\mathcal P_{ad}$. 

{ For $\tilde \tau\gg 1$, however,   Eq.~\eqref{smallw} describes only a
  small region close to $E_{ad}$ (see thin black lines 
in Fig.~\ref{pdfworktau}), and the overall shape depends both on 
$\alpha$ and $\tilde \tau$. 
{\g For $4\alpha\gg \tilde\tau$, almost all spectral weight is carried by the non-adiabatic processes ($\rho(w)$) around the typical value
$W_{typ}-E_{ad} \sim 2|E_{ad}|\ln(\tilde\tau)/\tilde\tau^2$, clearly separated from the adiabatic process.
 For $\tilde\tau\gg 4\alpha$, the adiabatic process gains spectral weight, $\mathcal P_{ad}\approx 1$, but a maximum for $W>E_{ad}$
remains present, though it gradually merges with the adiabatic processes.}
   
In particular, in the  small system limit} 
$\alpha_\tau \ll 1$, e.g., we can expand Eq.~\eqref{workgenquench} to get 
\begin{gather}
\rho(w)\approx \mathcal P_{ad}\alpha^2\left(\frac{\sin[w\alpha\tilde\tau/2]}{
  w\alpha\tilde\tau/2}\right)^2\exp(-\alpha w) 
\;.
\label{smallgamma}
\end{gather}
{Thus, for a linear quench, }
for energies commensurate with the quench time
the PDF is zero (see Fig. \ref{pdfworktau}). { This is related to}
the steady state behavior of the occupation numbers,   
$n_q\approx [{g_2(q)\sin(v|q|\tau)}/{2v^2|q|\tau}]^2 $
for $g_2\ll v$~\cite{EPAPS}, 
{ reflecting that} modes with energy commensurate to the quench time
stay { almost} unoccupied at $T=0$. 
{ In this limit ($\alpha\ll \tilde\tau $ and $\tilde \tau\gg 1$),  the system
 evolves almost adiabatically, non-adiabatic
 processes  have only a small probability $\sim \alpha/\tilde \tau$,   
and the typical work done in case of a rare non-adiabatic process is 
$W_{typ}\approx -\alpha^2\pi/\tau$.}


Increasing $\alpha$, the zeros of the PDF turn gradually 
into dips, { and the PDF develops a more universal form}. 
In the thermodynamic limit $\alpha_\tau \gg 1$,  using the
method  of steepest descent we obtain
\begin{gather}
\rho(w)\approx\frac{\mathcal P_{ad}\tilde \tau^{3/2}\sqrt\alpha}{2 \sqrt{\tan^3(s)\pi}}
\exp\left(w\left(\frac{\tilde \tau^2}{2}- \alpha\right)+2\frac{\alpha s}{\tilde\tau}\right)
\label{pdfsteepest}
\end{gather}
{\g for $\alpha s\gg \tilde\tau$,   with  $s\equiv\arctan[\sqrt{\exp( w\tilde \tau^2)-1}]$.
For $w \gg 1/\tilde\tau^2\gg1/\alpha^2$,  $\rho(w)$ in Eq. \eqref{pdfsteepest}
 behaves as a generalized Gumbel distribution of index 
$a=\frac{1}{2}+\frac{2\alpha}{\tilde\tau^2}$~\cite{clusel}.
 This latter emerges in the context of global fluctuations, describing the
 limit distribution of the $a$-th maximum of a sequence of independent
 and identically distributed random  
variables~\cite{gritsev}.
The distribution in the $1/\tilde\tau^2\gg w \gg 1/\alpha^2$ region resembles closely to Eq.~\eqref{gaussSQ}
apart from its normalization. 
}

{\em Experimental relevance.}
Our results can be tested on one-dimensional hard-core bosons~\cite{kinoshitascience} or non-interacting fermions as initial states.
The detection of the PDF of work requires two energy measurements, one before and one after
the time dependent protocol.
The first energy measurement can be omitted if we prepare the initial wave function in an energy eigenstate of $H(t=0)$.
The resulting energy distribution can then be probed
using time-of-flight experiments~\cite{polkovnikovrmp,BlochDalibardZwerger_RMP08}, similarly to  Ref. ~\cite{chen}.
The crossover between the various regimes can be monitored by tuning $\tau/\tau_0$ and $\alpha \sim N\left({g_2}/{v}\right)^2$, 
where $N$ is the number of particles in a 1D trap, typically with $N\sim 10^2$ - $10^3$ atoms~\cite{paredes,hofferberth,kinoshita}. 
By choosing $g_2/v\sim 1/\sqrt N$, $\alpha$ becomes of order unity, facilitating
the observation of crossover between the various regimes. For one-dimensional  interacting bosons (i.e. Bose-Hubbard model), 
$v\sim J$ and $g_2 \sim J^2/U$ for $U\gg J$ (close to the hard-core boson limit) with $U$ the on-site interaction~\cite{cazalillarmp} and $J$ the hopping amplitude.
By quenching away from the initial $U\gg J$ $\Leftrightarrow$ $g_2\approx 0$ limit (e.g. by changing the lattice parameters or tuning the Feshbach resonance),
a final interaction $U\sim J\sqrt N$ is reachable.
For weakly interacting fermions, $v\sim J$ and $g_2 \sim U$, therefore
ramping from the {\g weakly }interacting case to $U\sim J/\sqrt{N}$ is
desirable. {\g Nonetheless, our results apply also to interacting initial states\cite{EPAPS}.}

{\em Summary.}
We have studied the PDF of work done on a LL after an interaction
quench, realizable in strongly interacting Bose systems. 
We have constructed the density matrix of the generalized Gibbs ensemble with 
intermode correlations, describing arbitrary correlations of the steady state, thus the PDF of work.
The PDF exhibits markedly different characteristics depending on the
system size, 
 quench duration and interaction strength.
Our method is applicable to the full PDF of other observables as well, e.g. density fluctuations~\cite{armijo}.
We also emphasize that our results in Eqs.~\eqref{charworknq} and
\eqref{rhoGGE}  apply  also  
to a variety of other systems with effective bosonic Hamiltonians as in 
Eq. \eqref{hamilton}, including  interacting higher dimensional bosons
or  spin systems within a spin-wave theory. 

\begin{acknowledgments}
This research has been  supported by the Hungarian Scientific 
Research Funds Nos.  K72613, K73361, 	
K101244, CNK80991,
T\'{A}MOP-4.2.1/B-09/1/KMR-2010-0002  
and by the Bolyai program of the  Hungarian Academy of Sciences.
\end{acknowledgments}

\bibliographystyle{apsrev}
\bibliography{wboson}

\newpage

\section{Supplementary material for "Generalized Gibbs ensemble and  work statistics of a quenched Luttinger liquid"}

\setcounter{equation}{0}
\renewcommand{\theequation}{S\arabic{equation}}

\setcounter{figure}{0}
\renewcommand{\thefigure}{S\arabic{figure}}

\section{Quenching between interacting Luttinger liquids}

We demonstrate here that our results for the PDF of work applies also when 
we quench between interacting initial and final states. More precisely, we do not need to start directly
from the hard core boson limit for a bosonic LL or from strictly non-interacting fermions for a fermionic LL.
Let's consider an initially interacting LL, given by
\begin{equation}
H_i=\sum_{q\neq 0} \omega_q b_q^+ b_q
+\frac{g_q^i}{2}[b_qb_{-q}+b_q^+b_{-q}^+]\;,
\end{equation}
where $g_q^i$ is the initial interaction.
This can conveniently be diagonalized by a standard, time independent Bogoliubov transformation as
\begin{gather}
b_q=\cosh(\phi_q)a_q-\sinh(\phi_q)a_{-q}^+,\label{bt}\\
\tanh(2\phi_q)=\frac{g_q^i}{\omega_q},
\end{gather}
yielding
\begin{gather}
H_i=E_i +\sum_{q\neq 0} \tilde \omega_q a_q^+ a_q,
\end{gather}
where $\tilde \omega_q=\sqrt{(\omega_q)^2-(g_q^i)^2}$, and $E_i$ is the ground state energy of $H_i$ with respect to the non-interacting ground states.
The ground state of this Hamiltonian is the vacuum of the $a$ bosons.

The interaction quench is described, in the language of the initial bosonic operators,  by the additional term
\begin{gather}
H'=\sum_{q\neq 0}\frac{\Delta g_q(t)}{2}[b_qb_{-q}+b_q^+b_{-q}^+],
\end{gather}
where $\Delta g_q(t)=(g_q^f-g_q^i)Q(t)$, and $g_q^f$ is the final interaction strength.
After applying Eq. \eqref{bt} to this, we obtain the total time dependent Hamiltonian, $H=H_i+H'$ as
\begin{gather}
H=E_i+\sum_{q\neq 0}\left\{\left( \tilde \omega_q-\frac{\Delta g_q(t)}{2}\frac{g_q^i}{\tilde\omega_q}\right)  a_q^+ a_q+\right.\nonumber\\
\left.+\frac{\Delta g_q(t)}{2}\frac{\omega_q}{\tilde\omega_q}[a_qa_{-q}+a_q^+a_{-q}^+]-\frac{\Delta g_q(t)}{2}\frac{g_q^i}{\tilde\omega_q}\right\}.
\label{hamfinal}
\end{gather}
This, apart from the constant first and last terms on the r.h.s, is identical to Eq. (1) in the main text, after redefining its parameters, $v$, $\delta v$ and $g_q(t)$ 
 appropriately.
Therefore, our results for the PDF of work apply for any initial  and final interactions, provided that we stay in the perturbative regime throughout the quench, namely 
$\omega_q\gg |g_q^i|, |g_q^f|$.
The threshold above which work is possible, occurs at the difference of the adiabatic ground
state energies of the final and initial  states, namely at $E_{ad}=-({L/v})[(g_2^f)^2-(g_2^i)^2]/(v\tau_0)^2 16 \pi +\dots$. The orthogonality exponent
is determined by another energy scale, $E_{oe}=|E_{ad}(g_2^f-g_2^i)/(g_2^f+g_2^i)|$, giving $\alpha=E_{oe}\tau_0$.
After redefining $w=(W-E_{ad})/E_{oe}$, our results for $p(w)$ hold for this general case as well.

\section{Calculating the characteristic function of work}

The Hamiltonian in the Heisenberg picture for $t>\tau$ is given by
\begin{gather}
H_H(t)=\sum_{q>0}c_0(q,t)\left(b_q^+ b_q+b_{-q} b^+_{-q}\right)-\omega_q(t)+\nonumber\\
+c_1(q,t)b_qb_{-q}+c_1^*(q,t)b_q^+b_{-q}^+.
\end{gather}
where
\begin{gather}
c_0(q,t)=\omega_q(t)\left(|u_q(t)|^2+|v_q(t)|^2\right)+\nonumber\\
+g_q(t)2\textmd{Re}[u_q(t)v_q^*(t)],\\
c_1(q,t)=2\omega_q(t)u_q(t)v_q(t)+g_q(t)(u_q^2(t)+v_q^2(t)).
\end{gather}
This is simplified upon realizing that the operators
\begin{gather}
K_0(q)=\frac{b^+_qb_{q}+b_{-q}b^+_{-q}}{2}, \\
K_+(q)=b^+_qb^+_{-q}, \hspace*{1cm} K_-(q)=b_qb_{-q}
\end{gather}
 are the generators of a
SU(1,1) Lie algebra, satisfying $[K_+(q),K_-(q)]=-2K_0(q)$, $[K_0(q),K_\pm(q)]=\pm K_\pm(q)$, and the operators
for distinct $q$'s commute with each other.
Following Ref. \cite{truax}, we obtain
\begin{gather}
\exp[i \lambda H_H(t)]=\prod_{q> 0}\exp[\beta_+(q,\lambda,t) K_+(q)-i\omega_q(t)\lambda]\times\nonumber\\
\times\exp[2\beta_0(q,\lambda,t) K_0(q)]\exp[\beta_-(q,\lambda,t)K_-(q)],
\end{gather}
and the $\beta_{\pm,0}(q,\lambda,t)$ coefficients can be determined using Ref. \cite{truax}.
When starting from the ground state at $T=0$, the characteristic function of work is obtained as
\begin{gather}
\tilde G(\lambda,\tau)=\exp\left(\sum_{q>0}\beta_0(q,\lambda,t)-i\omega_q(t)\lambda\right)
\label{charwork}
\end{gather}
with
\begin{gather}
\beta_0(q,\lambda,t)=-\ln\left(\cos(\Omega_q\lambda)-\right.\nonumber\\
\left.-i\frac{\omega_q(t)+2\textmd{Im}[v_q^*(t)\partial_t v_q(t)]}{\Omega_q}\sin(\Omega_q\lambda)\right),
\label{beta0}
\end{gather}
where $\Omega_q=\sqrt{\omega_q^2(t>\tau)-g_q^2(t>\tau)}$ is the time independent adiabatic eigenenergy of a given mode in the final LL state\cite{giamarchi}.
After some trivial steps, Eq. \eqref{charwork} is rewritten as 
\begin{gather}
\ln\left(\tilde G(\lambda,\tau)\right)=-\sum_{q>0}\ln\left(1+n_q(1-\exp[2i\Omega_q\lambda])\right)+\nonumber\\
+i\lambda E_{ad},
\end{gather}
where
\begin{gather}
n_q=\frac{\omega_q(t)+2\textmd{Im}[v_q^*(t)\partial_t v_q(t)]}{2\Omega_q}-\frac 12
\label{nq}
\end{gather}
is the occupation number in the steady state
and
\begin{gather}
E_{ad}=\sum_{q>0}\left[\Omega_q-\omega_q(t>\tau)\right].
\end{gather}
is time independent, and stands for the difference of the adiabatic ground state energies of the final and initial states.

\begin{figure}[h!]
\psfrag{x}[t][][1][0]{$\omega_{av}(q)\tau/2\pi$}
\psfrag{y}[b][t][1][0]{$(v/g_2)^2 n_q$}
\includegraphics[width=5.5cm]{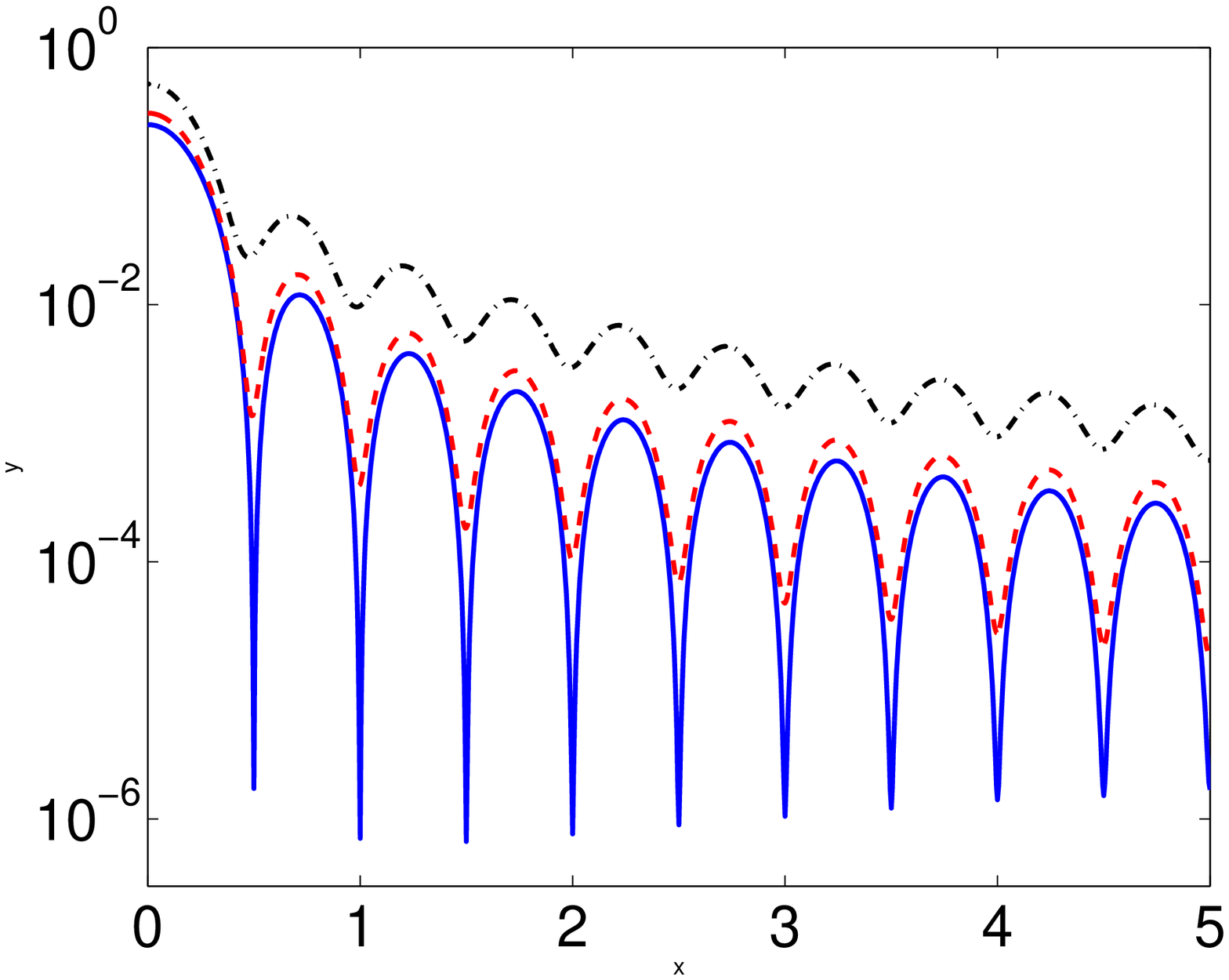}
\caption{The numerically evaluated occupation numbers in the steady state are plotted on a semilogarithmic scale 
after a linear quench with $\delta v=0$, 
$g_2/v=0.1$ (blue solid line), 0.5 (red dashed line) and 0.8  (black dash-dotted line), ranging from weak to  strong interactions.
The result of Eq. \eqref{nqapprox} is practically indistinguishable from the blue solid curve except close to 
$\omega_{av}(q)=2\pi n/\tau$ with $n$ integer, where the effective temperatures are the lowest.
The figure is valid for arbitrary $q$ since it depends only on the dimensionless combination $\omega_{av}(q)\tau$.}
\label{nqexact}
\end{figure}

The occupation number in the steady state is obtained for a linear quench from Eq. \eqref{nq} using the results of Ref. \cite{doraquench}
as
\begin{gather}
n_q\approx \left[\frac{g_2(q)\sin(v|q|\tau)}{2v^2|q|\tau}\right]^2
\label{nqapprox}
\end{gather}
for $g_2\ll v$. We have checked this prediction by numerically solving the differential equation in Eq. (3) in the main text.
The occupation numbers turn out to be periodic in the time averaged mode energies, defined by
\begin{gather}
\omega_{av}(q)=\frac{1}{\tau}\int\limits_0^\tau \sqrt{\omega_q^2(t)-g_q^2(t)}dt,
\end{gather}
and the numerically evaluated occupation numbers are plotted in Fig. \ref{nqexact}.
States with average energy commensurate with $\tau$ possess the lowest effective temperatures. 
The zeros predicted by Eq. \eqref{nqapprox} turn to sharp dips with increasing interaction, and our analytical expression describes rather reliably the numerical data.
Although for finite $g_2$, all effective temperatures are finite,
the steady state is far from being thermal, as evidenced by the highly non-thermal structure of the steady state density matrix.

\section{On the steady state density matrix from GGE}

The steady state density matrix reveals a highly non-thermal structure, namely
\begin{gather}
\hat\rho_{G}=\frac{1}{Z_{G}}\prod_{q>0}\exp\left[-\beta_q\Omega_q\hat n_q\right]\delta_{\hat n_q,\hat n_{-q}},
\label{dmss}
\end{gather}
possessing finite matrix elements only along $\hat n_q=\hat n_{-q}$. Thermalization would mean the same matrix elements for a fixed $\hat n_q+\hat n_{-q}$, which is not the case here.
Therefore, one can ask to what extent this density matrix is immune to additional perturbations, not considered within our Hamiltonian.

On the one hand, one can argue that experimental results on one dimensional cold atoms did not find any sign of thermalization, but 
rather prethermalization, i.e. reaching a certain non-thermal
steady state, took place\cite{kinoshita,gring,hofferberth,hofferberthnatphys}. These experimental results are nicely accounted for 
by a simple Gaussian model, similar to our Eq. (1) in the main text,
without additional terms.
On the other hand, from a theoretical point of view, additional interactions between the $b$ bosons are generated by the non-linearity of the non-interacting dispersion relation.
As was investigated in Ref. \cite{samokhin,imambekovrmp}, this broadens the otherwise sharp peak around 
the bare dispersion at $\omega_q=v|q|$ in the spectral function of the bosons, 
which, as a rough estimate, scales with $ \sim q^2/m$, 
where $m$ is the effective mass arising from curvature effects.
In the long wavelength limit ($q\sim 0$), this gives a "lifetime", which is usually much longer that the typical experimental timescales.
Higher order terms in the $b$ bosons also arise from density-density  interactions\cite{cazalillarmp}, 
but these provide higher powers of $q$ in the broadening of the bosonic mode, 
suppressing further their effect.

Nonlinear terms of sine-Gordon type \cite{giamarchi} are absent without any lattice, i.e. for interacting 
particles in the continuum limit, which can also be realized experimentally\cite{kinoshitascience}.
In the presence of an optical lattice, these are inevitably present, though their effect can be weakened by
choosing incommensurate fillings or suppressing spin backscattering (e.g. by using single component bosons).
Therefore, our non-thermal density matrix in Eq. \eqref{dmss} is expected to describe fairly reliably the steady state of interaction quenched one dimensional systems.

Having established the validity of our density matrix, we now turn to the derivation of results, presented in the main text, using the steady state density matrix.
The partition function of the GGE is determined as
\begin{gather}
Z_G=\prod_{q>0}\frac{1}{1-\exp[-\beta_q\Omega_q]},
\end{gather}
and the characteristic function of work reads as
\begin{gather}
\tilde G(\lambda, \tau)\exp(-i\lambda E_{ad})=\nonumber \\
=\frac{1}{Z_G}\prod_{q>0}\sum\limits_{n_q=0}^\infty \exp\left[-\beta_q\Omega_q+2i\lambda\Omega_qn_q\right]=\nonumber\\
=\prod_{q>0}\frac{1-\exp[-\beta_q\Omega_q]}{1-\exp[-\beta_q\Omega_q+2i\lambda\Omega_q]}=\nonumber\\
=\exp\left(-\sum_{q>0}\ln\left(\frac{\exp[\beta_q\Omega_q]-\exp[2i\lambda\Omega_q]}{\exp[\beta_q\Omega_q]-1}\right)\right)=\nonumber\\
=\exp\left(-\sum_{q>0}\ln\left(1+\frac{1-\exp[2i\Omega_q\lambda]}{\exp(\beta_q\Omega_q)-1}\right)\right).
\end{gather}

\section{The cumulants of energy}

\begin{figure}[h!]
\includegraphics[width=5.5cm]{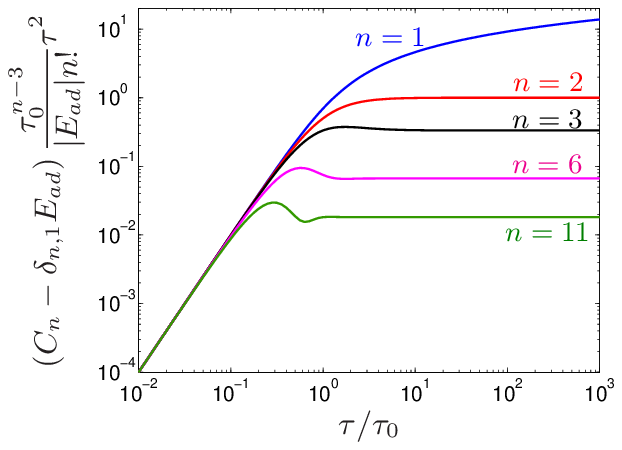}
\caption{(Color online) Several cumulants of the work done on a LL are log-log plotted as a function of the quench time for a linear protocol.
Close to the SQ limit ($\tau\ll\tau_0$), all properly normalized cumulants are equal, while in the near adiabatic limit ($\tau\gg\tau_0$), these approach $2/n(n-1)$.}
\label{cumulantslinearquench}
\end{figure}

The cumulants, $C_n$ of the PDF of work done
can be derived after expanding the characteristic function of work done 
in power series as $\ln\left(\tilde G(\lambda,\tau)\right)=\sum_{n=1}^\infty C_n (i\lambda)^n/n!$, yielding
\begin{gather}
\frac{C_{n}}{E_{ad}}=\delta_{n,1} -\int\limits_0^\tau\int\limits_0^\tau \frac{Q'(t_1)Q'(t_2)\tau_0^2 n!}{\left[i(t_1-t_2)-\tau_0\right]^{n+1}}dt_1 dt_2,
\label{workcumulants}
\end{gather}
where the first term denotes the  adiabatic ground state energy difference between the initial and final state, while the second one stems from the non-adiabatic
evolution (i.e. heat).
 The first cumulant, $C_1=\langle H(t>\tau)\rangle$ was already calculated in Ref. \onlinecite{doraquench}.
Their behaviour is illustrated in Fig. \ref{cumulantslinearquench} after a linear quench: $C_1-E_{ad}$ decays as $\ln(\tau/\tau_0)/\tau^2$, since $Q'(t)$ exhibits
kinks at $t=0$ and $\tau$\cite{doraquench,degrandi}, consequently all $C_{n>1}$ decay as $\tau^{-2}$.

For a SQ, the cumulants are $C_1=\langle H(t>0)\rangle=E_{SQ}(=0$ to second  order in $g_2$ and $\delta v$ within our scheme), $C_{n>1}=\alpha n!/\tau_0^{n-2}$.

\section{Analytical results for a linear quench}

In the case of a linear quench, the characteristic function of work is obtained as
\begin{gather}
\ln\left[\tilde G(\lambda,\tau)\right]=iE_{ad}\left[\lambda -\frac{\tau_0^2}{\tau^2}\left(g(\lambda+i\tau_0)-g(i\tau_0)\right)\right],
\label{charworklinear}
\end{gather}
where $g(z)=(z+\tau)\ln(z+\tau)+(z-\tau)\ln(z-\tau)-2z\ln(z)$.
The  cumulants are obtained from  a variant of Eq. \eqref{workcumulants} as
\begin{gather}
\frac{C_{n}}{E_{ad}}=\delta_{n,1} -{(-1)^n \tau_0^2}\frac{\partial^{n-1}}{\partial \tau_0^{n-1}}\frac{\Delta E}{E_{ad}\tau_0^2},
\label{ddd}
\end{gather}
where the heating,  $\Delta E$ is
\begin{gather}
\Delta E=-E_{ad}\frac{\tau_0^2}{\tau^2}\ln\left(1+\frac{\tau^2}{\tau_0^2}\right),
\end{gather}
and Eq. \eqref{ddd} holds true for an arbitrary quench protocol.
The asymptotic behaviour of Eq. \eqref{charworklinear} for $\lambda\rightarrow\infty$ gives
\begin{gather}
\tilde G(\lambda\rightarrow\infty,\tau)=\exp\left(i\lambda E_{ad}-2\frac{\alpha }{\tilde \tau}\arctan\left(\tilde\tau\right)\right)\times\nonumber\\
\times
\left(1+\tilde \tau^2\right)^{\alpha/\tilde\tau^2}\hspace*{-4mm}=\left\{
\begin{array}{cc}
\exp\left(i\lambda E_{ad}-\dfrac{\pi \alpha}{\tilde \tau}\right) & \tau\rightarrow \infty,\\
\exp\left(i\lambda E_{ad}-\alpha\right) & \tau\rightarrow 0.
\end{array}\right.
\end{gather}
From this, the asymptotic behaviour of $P_{ad}$ follows as
\begin{gather}
P_{ad}=\left\{
\begin{array}{cc}
\exp\left(-\alpha+\dfrac{\alpha\tilde\tau^2}{6}\right), & \tau\rightarrow 0,\\
1-\dfrac{\pi\alpha}{\tilde\tau},& \tau\rightarrow \infty.
\end{array}
\right.
\end{gather}

\end{document}